\documentclass[preprint]{ptephy_v2}%%%%%% to generate preprint number

\preprintnumber{XXXX-XXXX} %%% %%% Insert preprint number here
\usepackage{hyperref}
\usepackage{float}
\newcommand{\be}{\begin{equation}}
\newcommand{\ee}{\end{equation}}

\begin{document}

\title{Constant-roll $\beta$-exponential inflation: Palatini formalism}

%%%% To generate auto affiliation numbers please use \author{}\affil{} command

\author{Ozan Sarg{\i}n}
\affil{Sabanc{\i} University, \\ Faculty of Engineering and Natural Sciences, \\ 34956 Tuzla, {\.I}stanbul, T\"{u}rkiye \email{ozansrgn@gmail.com}}

%\author{Insert second author name here}
%\affil{Insert second author address here}

%\author{Insert third author name here}
%\author[3]{Insert fourth author name here} %%% Use optional bracket [3] to change the respective address
%\affil{Insert third author address here}

%\author{Insert last author name here\thanks{These authors contributed equally to this work}}
%\affil{Insert last author address here}

%%% To include the collaborator name... Please use the command "\collaborator"
%%% For example: \collaborator{ATLAS Collaboration}

\begin{abstract}
This paper investigates the inflationary dynamics of a $\beta$-exponential potential model within the framework of non-minimally coupled quadratic $(R+R^2)$ gravity. The functional form of the adopted potential provides a well-motivated framework; its physical origin can be interpreted either as the stabilization dynamics of a radion field determining the size of the extra dimension in braneworld cosmology, or as a manifestation of the $q$-exponential function emerging naturally from Tsallis non-extensive thermodynamics. Utilizing the Palatini formalism, we derive an effective Einstein-frame generalized k-inflation theory and analyze its evolution under the constant-roll condition. We perform a comprehensive scan of the parameter space to obtain predictions for the spectral index $n_s$ and the tensor-to-scalar ratio $r$. Our results demonstrate that for specific viable ranges of the model parameters, the inflationary observables are in excellent agreement with the latest observational data from the Atacama Cosmology Telescope ({\textsf{ACT}}) DR6 and the Planck mission, thereby identifying the physically consistent regions of the parameter space. Furthermore, an evaluation of the primordial non-Gaussianity confirms that the constant-roll dynamics generate a distinct, observationally viable phenomenological signature.
\end{abstract}

\subjectindex{}

\maketitle

%%%%%%%%%%%%%%%%%%%%%%%%%%%%%%%%%%%%%%%%%%%%%%%%%%%%%%%%%%%%%%%%%%%%%%%%%%%%%%%%%%%%%%%%%%
\section{Introduction} \label{sec:intro}

The theory of cosmic inflation posits a phase of quasi-de Sitter expansion in the early Universe. This framework was developed to address critical issues in traditional Big Bang cosmology, notably the observed flatness and uniformity of the Cosmic Microwave Background (CMB) temperature across vast scales \cite{Starobinsky:1980te, Guth1981, Sato1981, Linde1982, Albrecht1982, Linde1983a, Lyth1999}. The significance of inflation has evolved, offering a lens through which the dynamics of minute primordial inhomogeneities originating from quantum fluctuations can be understood as they scale up beyond the causal horizon \cite{Starobinsky1979, Mukhanov1981b, Hawking1982a, Hawking1983, Starobinsky1982a, Guth1982a}. Recent advancements in CMB observational techniques \cite{Akrami2018, BICEP:2021xfz, ACTcollab2} are increasingly constraining the inflationary spectrum, thereby narrowing the viable parameter space for inflationary models. Although the scalar perturbation power spectrum remains approximately scale-invariant and Gaussian, the potential for identifying non-Gaussian signatures is increasing, thanks to these observational improvements.

In this inflationary context, the scalar degree of freedom known as the inflaton can be viewed as either a fundamental scalar field or as an effective scalar degree of freedom stemming from gravitational theories themselves. This latter interpretation is particularly relevant in modified gravity scenarios like $f(R)$ gravity \cite{DeFelice2010, Sotiriou2010, Capozziello2011, Clifton2012, Odintsov2016, Bostan_2024}, exemplified by the Starobinsky model, which is consistent with observational data regarding its inflationary spectrum. The Starobinsky model can be recast as a scalar-tensor theory distinguished by non-minimal coupling between the effective scalar field and the Ricci scalar, as is the case for all gravitational theories expressed via an $f(R)$ action.

The Palatini formulation of gravity \cite{Palatini:1919}, though historically acknowledged, has seen a resurgence in interest \cite{Bauer2008, Borunda2008, Bauer2011, Tamanini2011, Enqvist2012, Borowiec2012, Racioppi2017, Rasanen2017, Fu2017, Stachowski2017, Azri2017, Enckell2018, Enckell2018a, Jaerv2018, Rasanen2018, Antoniadis2018, Rasanen2018a, Antoniadis2019b, Racioppi2019, Shimada2019, Jinno2019, Tenkanen2019, Rubio2019, Gialamas2019, Tenkanen2020, Shaposhnikov2020, Tenkanen2020a, Shaposhnikov2020a, Lloyd-Stubbs2020, Gialamas_2021r, Gialamas2020b, Bostan_2021r, Bekov2020a, Dimopoulos2020a, Gialamas2020a, Annala_2021a, Dioguardi_2022a, Bostan_2023a, Dioguardi_2024a, EADKHONG2023116289, Gialamas2023a, Laur2025, Hassan_2025, Bostan_2025d, Bombacigno_2025, Bostan_2025b, Dimopoulos2025, PALLIS2025139739, Bostan2026} due to its unique predictions relative to the standard metric formulation, particularly in the context of non-minimally coupled scalar fields. In the Palatini approach, both the metric $g_{\mu\nu}$ and the connection  $\Gamma_{\mu\nu}^\rho$  are treated independently. While both the Palatini and metric formulations yield equivalent results within General Relativity, they diverge significantly in theories that incorporate non-minimal couplings of the inflaton to gravity. For instance, when analyzing the Starobinsky model in the Palatini context, one finds a lack of propagating scalar degrees of freedom, contrasting sharply with the metric formulation.

By integrating a quadratic gravity model featuring an $R^2$ term, along with a non-minimally coupled fundamental scalar expressed as $\xi\,\phi^2\,R$ within the Palatini framework, one establishes a generalized $k$-inflation model \cite{Armendariz-Picon1999, Chiba2000, Armendariz-Picon2000, Armendariz-Picon2001, Chiba2002, Malquarti2003, Malquarti2003a, Chimento2004, Chimento2004a, Scherrer2004, Aguirregabiria2004, Armendariz-Picon2005, Abramo2006, Rendall2006, Bruneton2007, Putter2007, Babichev2008, Matsumoto2010, Deffayet2011, Unnikrishnan2012, Li2012, Serish2025, Bilic_2025, Bilic_2025a, LIU2025102039, Ferreira2024a, Ageeva2024a, YANG2024101560, Shen2023a, Nguyen2022, Shumaylov2022a, Mikura2021a, Pareek2021a, Solbi_2021a, Oikonomou2021a} that incorporates both field-dependent quartic and quadratic kinetic terms.
This paper explores the phenomenology of these models by moving beyond the conventional slow-roll approximation, which typically overlooks the influence of the inflaton's second derivative in its equations of motion. We propose that a constant-roll condition holds, such that $\ddot{\phi}/H\dot{\phi}$ remains approximately constant \cite{Motohashi2015a, Motohashi2017a, Odintsov2017, Odintsov2017a, Nojiri2017, Motohashi2017b, Gao2017a, Oikonomou2017, Odintsov2017c, Oikonomou2017a, Cicciarella2018a, Anguelova2017, Karam2018c, Yi2017, Gao2019, Odintsov2020, KESKIN2025139947, Inui_2025c, Biswas2024a, HERRERA2024169705, Mohammad_Ahmadi_2024, Tomberg2023a, NOJIRI2023137926, Motohashi_2023q, Mohammadi_2023q, Panda2023q, Ravanpak2022q, Anari_2022q, Setare2021q, AlHallak2022, Garnica2022, Shokri2022q, Nguyen2021q, SHOKRI2022100923, Shokri_2022q, Oikonomou2022q, Mun2021q, SHOKRI2021168487, Shokri2021qa, Sadeghi2021aq, Guerrero2020q, Gao2020aq, Oikonomou_2020aqw, Mohammadi2020aq}. A feature of the constant-roll paradigm is its correlation with non-Gaussianities in the CMB spectrum, contrasting with the minimal non-Gaussian effects anticipated under the slow-roll approximation for $k$-inflation models.

In this paper, we investigate a scalar field model coupled to gravity, incorporating an $R^2$ term within the Palatini formalism. We focus on the inflationary dynamics under the constant-roll condition, specifically examining a $\beta$-exponential potential and the implications of the non-minimal coupling of the inflaton to gravity. 

The $\beta$-exponential potential serves as a generalization of conventional power-law inflation, as introduced in \cite{Alcaniz:2006nu}. Ref. \cite{Alcaniz:2006nu} explores the inflationary predictions related to a minimally coupled scalar field with the generalized $\beta$-exponential potential, motivated by the prevalence of scalar fields with exponential potentials in various particle physics theories. Notably, this includes models from supergravity and string theory—such as the well-known Salam-Sezgin model \cite{halliwell}—as well as gravitational theories involving high derivative terms \cite{wett,wett1} and Kaluza-Klein scenarios where extra dimensions are compactified. The findings in \cite{Alcaniz:2006nu} suggest that this single scalar field framework enables a broader spectrum of solutions compared to standard exponential models, yielding viable predictions for the scalar spectral index and the tensor-to-scalar ratio.

Subsequent investigations, such as \cite{Santos:2017alg} and \cite{dosSantos:2021vis}, delve into CMB constraints on this model. The first paper emphasizes the minimal coupling scenario for the inflaton, while the latter integrates non-minimal coupling, refining the analysis with updated data from Planck (2018) \cite{Akrami2018}. A distinctive characteristic of the $\beta$-exponential potential is its deep roots in fundamental physics; it not only derives from braneworld cosmological frameworks—where the radion determining the extra-dimensional size acts as the inflaton \cite{Santos:2017alg}—but also emerges naturally from Tsallis non-extensive thermodynamics as a manifestation of the $q$-exponential function \cite{Tsallis1988, Kumar2022, Sheykhi2018, Nojiri2019}.

While the aforementioned studies, \cite{Santos:2017alg} and \cite{dosSantos:2021vis}, predominantly focus on standard slow-roll inflation, a notable exception is \cite{Santos:2022exm}, which explores warm inflation dynamics for this potential, where inflaton decay to radiation occurs during inflation. The relatively recent work on the $\beta$-exponential potential, \cite{bostan} examines inflationary observables in the Palatini formulation for a minimally coupled inflaton across various reheating temperatures.

The relevance of the $\beta$-exponential potential has been further highlighted by very recent studies addressing the emerging discrepancies between the {\textsf{Planck}} 2018 results and the latest {\textsf{ACT}} alongside Dark Energy Spectroscopic Instrument ({\textsf{DESI}}) data. The combined {\textsf{ACT}} and {\textsf{DESI}} observations indicate a preference for a bluer scalar spectral index ($n_s \approx 0.974$), placing significant tension on standard universal attractor models. Very recently, Ref. \cite{Yuennan2026} demonstrated that both minimally and non-minimally coupled $\beta$-exponential models can naturally resolve this $n_s$ tension. Their perturbative slow-roll analysis confirmed that this potential accommodates the higher $n_s$ values favored by {\textsf{ACT}}. However, their predicted tensor-to-scalar ratio remains relatively large ($r \sim \mathcal{O}(10^{-2})$) and strictly requires significantly large deformation parameters ($\beta \sim 5.0$) to safely satisfy the current {\textsf{BICEP/Keck}} upper bounds.

Building upon these compelling and timely findings, in this paper, we also investigate the observational indices for the $\beta$-exponential potential. However, to extend the dynamics beyond the conventional slow-roll regime and achieve a rigorous suppression of the tensor modes naturally, we impose a constant-roll condition on the rate of roll of the inflaton within the Palatini quadratic gravity framework. Our findings reveal that the Einstein frame Lagrangian corresponds to a generalized $k$-inflation model, with predicted observables aligning with the constraints imposed by the recent sixth data release (DR6) of the Atacama Cosmology Telescope ({\textsf{ACT}}) collaboration \cite{ACTcollab2} over a reasonable range of model parameters.

The paper is structured as follows: we begin, in \S \ref{Sec.2}, by formulating the $f(R,\phi)$ gravity theory utilizing an auxiliary field $\chi$, while considering the presence of a fundamental scalar field $\phi$ that exhibits non-minimal coupling and possesses a potential $V(\phi)$. Through a Weyl rescaling of the metric and the subsequent resolution of the auxiliary field's constraint equation, we derive the Einstein frame Lagrangian. Then, by employing a flat FRW background metric, we extract the equations of motion governing the theory.

In \S \ref{Sec.3}, we describe the inflationary parameters and observables, operating under the constant-roll condition. \S \ref{Sec.4} introduces the $\beta$-exponential model. In \S \ref{Sec.5}, alongside a comprehensive numerical analysis of the primordial tilt and the tensor-to-scalar ratio, we evaluate the equilateral non-Gaussianity amplitude to rigorously substantiate the physical significance of the constant-roll condition. We culminate our work in the final section with a summary of our conclusions.
%%%%%%%%%%%%%%%%%%%%%%%%%%%%%%%%%%%%%%%%%%%%%%%%%%%%%%%%%%%%%%%%%%%%%%%%%%%%%%%%%%%%%%%%%%

\section{\texorpdfstring{$f(R,\phi)$}{f(R,phi)} gravity in the Palatini formalism} \label{Sec.2}

We consider a Lagrangian with a scalar field $\phi$ that is non-minimally coupled to gravity, including a quadratic curvature term. The corresponding action in the Jordan frame is
\be\label{ac1}
    \mathcal{S}=\int\!\mathrm{d}^4x\,\sqrt{-g}\left\{\frac{1}{2}(1+\xi\phi^2)R+\frac{\alpha}{4}R^2-\frac{1}{2}(\nabla\phi)^2-V(\phi)\right\},
\ee
where $\xi$ and $\alpha$ are dimensionless parameters, and we employ natural units, i.e., we assume $M_P^2\equiv 1$.
The same theory can be expressed in the equivalent scalar-tensor representation as
\be\label{ac2}
    \mathcal{S}=\int\!\mathrm{d}^4x\,\sqrt{-g}\left\{\frac{1}{2}(1+\xi\phi^2+\alpha{\theta}\,^2)R-\frac{\alpha}{4}{\theta}^4-\frac{1}{2}(\nabla\phi)^2-V(\phi)\right\},
\ee
where $\theta$ is an auxiliary scalar that satisfies ${\theta}\,^2=R$ on-shell. After letting $1+\alpha {\theta}\,^2\!=\!\chi^2$, the action in \eqref{ac2} becomes
\be\label{ac3}
    \mathcal{S}=\int\!\mathrm{d}^4x\,\sqrt{-g}\left\{\frac{1}{2}(\chi^2+\xi\phi^2)R -\frac{(\chi^2 -1)^2}{4 \alpha} -\frac{1}{2}(\nabla\phi)^2-V(\phi)\right\}.
\ee
The theory can be conformally transformed into the Einstein frame via Weyl rescaling the metric through 
\be \label{ac4}
   \bar{g}_{\mu\nu}=(\chi^2+\xi\phi^2)\,g_{\mu\nu}.
\ee
After dropping the bars on the metric, we obtain
\be\label{acEF}
    \mathcal{S}=\int\!\mathrm{d}^4x\,\sqrt{-g}\left\{\frac{1}{2}R-\frac{1}{2}\,\frac{(\nabla\phi)^2}{(\chi^2+\xi\phi^2)}-\frac{4\alpha V(\phi) + (\chi^2-1)^2}{4\alpha (\chi^2+\xi\phi^2)^2}\right\}.
\ee
The key aspect of the Palatini formalism lies in the fact that the Ricci scalar \( R = g^{\mu\nu} R_{\mu\nu}(\Gamma) \) undergoes a rescaling during a conformal transformation. This implies that the scalar degree of freedom $\chi$ maintains its auxiliary role. As a result, we can eliminate the non-dynamical field $\chi$ from the action by solving the constraint equation $\delta_\chi \mathcal{S} = 0$ for $\chi^2$. The result is 
\be \label{aux}
    \chi^2=\frac{1+\xi\phi^2+4\alpha V(\phi)+\alpha\xi\phi^2(\nabla\phi)^2}{(1+\xi\phi^2)-\alpha(\nabla\phi)^2}\, .
\ee
Substituting \eqref{aux} back into \eqref{acEF}, we arrive at the effective action
\begin{equation}\label{efacc}
    \mathcal{S}=\int\!\mathrm{d}^4x\,\sqrt{-g}\left(\frac{1}{2}R+\mathcal{L}(\phi,X)\right),
\end{equation}
which is expressed in terms of the Lagrangian density 
\begin{equation}\label{lag1}
    \mathcal{L}(\phi,X) \equiv A(\phi)X+B(\phi)X^2-U(\phi)\,,
\end{equation}
that belongs to a class of generalized $k$-inflation effective theory \cite{Antoniadis:2020dfq}.  

The Lagrangian features a quartic kinetic term alongside a quadratic one, both with field-dependent coefficients $A(\phi)$ and $B(\phi)$. In equation \eqref{lag1}, $X$ represents the kinetic energy density
$X\,\equiv\,\frac{1}{2}(\nabla\phi)^2$. 
Also, the field-dependent coefficients $A(\phi)$, $B(\phi)$, and the effective potential $U(\phi)$ read

\be \label{abu}
 \left.\begin{array}{r}
A(\phi)\,\equiv\,-\left[(1+\xi\phi^2)+\frac{4\alpha V(\phi)}{(1+\xi\phi^2)}\right]^{-1}\\
\,\\
B(\phi)\,\equiv\,2\alpha\left[\,(1+\xi\phi^2)^2+4\alpha V(\phi)\right]^{-1}\,=\,-\frac{2\alpha A(\phi)}{(1+\xi\phi^2)}\\
\,\\
U(\phi)\,\equiv\,V(\phi)\,\left[\,(1+\xi\phi^2)^2+4\alpha V(\phi)\right]^{-1}\,=\,-\frac{ V(\phi) A(\phi)}{(1+\xi\phi^2)}\;.
\end{array}\quad \right\}
\ee
The energy-momentum tensor of the source field $\phi$  obtained by varying \eqref{efacc} with respect to the metric $g_{\mu\nu}$ reads
\begin{equation}
T_{ \mu\nu}\,=\,-(A+2BX)\left(\nabla_{ \mu}\phi\right)\left(\nabla_{ \nu}\phi\right)+g_{ \mu\nu}(AX+BX^2-U)\,.
\end{equation}
In a spatially flat Friedmann-Lemaître-Robertson-Walker (FLRW) background metric, assuming the scalar field is homogeneous in space and varies only with time, we can derive its corresponding energy density and pressure using $\rho=T_{00}$ and $T_{ij}=p\,g_{ij}=\mathcal{L}\,g_{ij}$, yielding   
\begin{align}
    \rho&=A(\phi)X+3B(\phi)X^2+U(\phi),\label{g00}\\
    p&=A(\phi)X+B(\phi)X^2-U(\phi).
\end{align}
Einstein's equations of motion are:
\be\label{eden}
    3H^2=\rho,
\ee
\be\label{pres}
 \dot{\rho}+3H(\rho+p)=0,
 \ee
where the dot denotes the derivative with respect to cosmic time $t$ and $H$ is the Hubble parameter. Combining \eqref{eden} and \eqref{pres} gives
\begin{equation}
    2\dot{H}+3H^2=-p.
\end{equation}
Manipulating Einstein's equations together with the definitions of energy density and pressure gives the scalar field equation of motion as
\be \label{eqnmot}
    \ddot{\phi}(A+6BX)+3H\dot{\phi}(A+2BX)-A'X-3B'X^2=U' 
\ee
where the derivative with respect to $\phi$ is represented using prime notation.
%%%%%%%%%%%%%%%%%%%%%%%%%%%%%%%%%%%%%%%%%%%%%%%%%%%%%%%%%%%%%%%%%%%%%%%%%%%%%%%%%%%%%%%%%%%%%
\section{Constant-roll \texorpdfstring{$k$}{k}-inflation dynamics} \label{Sec.3}

The scalar field $\phi$, which represents the sole dynamical scalar degree of freedom, is considered to be a candidate for the inflaton. We will assume that the condition for constant-roll is met, as expressed by the equation 
\be\label{const}
    \ddot{\phi}=\kappa H\dot{\phi},
\ee
where $\kappa$ is a dimensionless real parameter. 

It is important to note that the constant-roll condition \eqref{const} approaches the slow-roll condition in the limit of $\kappa \ll 1$, or it becomes the ultra slow-roll condition when $\kappa = -3$ exactly. 

The slow-roll parameters (SRPs) are defined as follows~\cite{Odintsov2020}:
\begin{equation}\label{srps}
    \epsilon_1=-\frac{\dot{H}}{H^2},\qquad\epsilon_2=-\frac{\ddot{\phi}}{H\dot{\phi}},\qquad\epsilon_3=\frac{\dot{F}}{2HF},\qquad\epsilon_4=\frac{\dot{E}}{2HE},
\end{equation}
where
\begin{equation}
    F=\frac{\partial \mathcal{L}}{\partial R}, \qquad E=-\frac{F}{2X}\left(X\,\frac{\partial\mathcal{L}}{\partial X}+2X^2\,\frac{\partial^2\mathcal{L}}{\partial X^2}\right).
\end{equation}
These parameters are introduced under the assumption that the condition $\dot{\phi}^2/2\ll U(\phi)$ holds during at least the initial stages of inflation. The second SRP is constrained by the constant-roll condition \eqref{const} to be $\epsilon_2=-\kappa$. Moreover, since we are in the Einstein frame where $F=1/2$, it follows that $\epsilon_3=0$. 

In terms of SRPs, we can express observable quantities such as the scalar spectral index (also known as the primordial tilt) $n_s$ and the tensor-to-scalar ratio $r$ as follows~\cite{Noh2001a, Hwang2002, Hwang2005}:

\begin{align} 
    n_s &= 1 - 2 \frac{2 \epsilon_1 - \epsilon_2 - \epsilon_3 + \epsilon_4}{1 - \epsilon_1}, \label{ind1}\\
    r &= 4 |\epsilon_1| C_s, \label{ind2}
\end{align}
where $C_s$ represents the wave speed that characterizes the propagation of primordial perturbations, given by the equation:

\begin{equation}
    C_s^2 = \frac{\mathcal{L}_X}{\mathcal{L}_X + 2X \mathcal{L}_{XX}} = \frac{A(\phi) - B(\phi) \dot{\phi}^2}{A(\phi) - 3B(\phi) \dot{\phi}^2} = \frac{1 + \xi \phi^2 + 2\alpha \dot{\phi}^2}{1 + \xi \phi^2 + 6\alpha \dot{\phi}^2}\;.
\end{equation}
It is important to note that the sound speed $C_s$ meets the condition $0 < {C_s}^2 < 1$. On the other hand, the scalar power spectrum, evaluated up to first order in the SRPs, is given by~\cite{Lyth1999}
\begin{equation}\label{powspec}
P_S\,\approx\,\frac{H^2}{8\pi^2\epsilon_1}\,=\,\frac{1}{72\pi^2}\frac{\left(AX+3BX^2+U\right)^2}{\left(AX+2BX^2\right)}\,,
\end{equation}
which should yield the observed value $P_S\approx 2.1\times 10^{-9}$~\cite{Akrami2018, BICEP:2021xfz} at horizon crossing. Similarly, the integral expression for the number of $e$-folds is given as
\begin{equation} \label{perturb2}
N\,=\,\int_{\phi_i}^{\phi_f}\frac{d\phi}{\dot{\phi}}H\,=\,\frac{1}{\sqrt{3}}\int_{\phi_i}^{\phi_f}\frac{d\phi}{\dot{\phi}}\sqrt{AX+3BX^2+U}\,.
\end{equation}

The inflationary analysis can be conducted once all the expressions for the slow-roll parameters (SRPs), observational indices, and the number of $e$-folds are expressed as a function of the field $\phi$. To achieve this, we first need to solve the equation of motion of the scalar field \eqref{eqnmot} to find an expression for $\dot{\phi}[\phi]$. Before that, however, we must derive an approximate expression for the Hubble parameter, $H$.
We derive an expression for $H$ in terms of $\dot{\phi}^2/2U$ by combining equations \eqref{g00} and \eqref{eden} as follows:
\begin{eqnarray}  \label{hubble}
    H &=& \sqrt{\frac{U}{3}\left(1 - A\left(\frac{\dot{\phi}^2}{2U}\right) + 3BU\left(\frac{\dot{\phi}^2}{2U}\right)^2\right)}\nonumber  \\ 
    H &\approx& \sqrt{\frac{U}{3}}\left[1 - \frac{A}{2}\left(\frac{\dot{\phi}^2}{2U}\right) + \frac{1}{2}\left(3BU - \frac{A^2}{4}\right)\left(\frac{\dot{\phi}^2}{2U}\right)^2\right] 
\end{eqnarray} 
where, in the last step, a Taylor expansion in the small parameter $\dot{\phi}^2/2U$ is performed.

The next step is to substitute the constant-roll condition \eqref{const} into the equation of motion \eqref{eqnmot} to derive a first-order differential equation in terms of $\dot{\phi}$:
\be \label{foqdot}
  4H\dot{\phi}A(\kappa +3)-12H \dot{\phi}^3 B(\kappa +1)+2A'\dot{\phi}^2-3B'\dot{\phi}^4-4U'=0 \: .
\ee
Substituting  \eqref{hubble} into \eqref{foqdot}, we obtain
\begin{equation} \label{td1}
    \sqrt{\frac{U}{3}}\,\left(3B(\kappa+1)+\frac{A^2}{4U}(\kappa+3)\right)\dot{\phi}^3-\frac{1}{2}A'\dot{\phi}^2-A\sqrt{\frac{U}{3}}(\kappa+3)\dot{\phi}+U'=0,
\end{equation}
where we kept only terms up to $\mathcal{O}(\dot{\phi}^3)$.
After making a change of variables
\be
    x\,\equiv\,\dot{\phi}-\frac{A'}{6\gamma}
\ee
the equation \eqref{td1}, which is a third-degree polynomial in terms of $\dot{\phi}$, can be expressed
as
\begin{equation}  \label{tdx}
    x^3+c_1\,x+c_0=0\,,
\end{equation}
where
\begin{align}
    \gamma&=\sqrt{\frac{U}{3}}\,\left(3B(\kappa+1)+\frac{A^2}{4U}(\kappa+3)\right),\\
    c_1&=-\frac{1}{\gamma}\left[(\kappa+3)A\sqrt{\frac{U}{3}}+\frac{(A')^2}{12\gamma}\right],\\
    c_0&=\frac{1}{\gamma}\left[U'-A(\kappa+3)\,\frac{A'}{6\gamma}\sqrt{\frac{U}{3}}-\frac{(A')^3}{108\gamma^2}\right] \: .
\end{align}
The real-valued solution to \eqref{tdx} is
\be\label{solx}
    x=\frac{\left(-9c_0+\sqrt{3}\sqrt{4{c_1}^3+27{c_0}^2}\right)^{1/3}}{2^{1/3}\,3^{2/3}}-\frac{\left(\frac{2}{3}\right)^{1/3}c_1}{\left(-9c_0+\sqrt{3}\sqrt{4{c_1}^3+27{c_0}^2}\right)^{1/3}} \:\: .
\ee

The scalar spectral index, tensor-to-scalar ratio, and scalar power spectrum are all functions of the slow roll parameters, as indicated in equations \eqref{ind1}, \eqref{ind2}, and \eqref{powspec}. Therefore, we need to solve for the slow-roll parameters first.  Using the definitions in  \eqref{srps}, 
the SRPs become
\be\label{eps1}
\epsilon_{1}=\frac{3 A X +6 B X^2}{A X +3 B X^2+U} \:\, ,
\ee
\be\label{eps2}
\epsilon_{2}=-\kappa \, ,
\ee
\be\label{eps3}
\epsilon_{3}=0  \, ,
\ee
\be\label{eps4}
\epsilon_{4}= \frac{\sqrt{3}}{2}\,\frac{\dot{\phi}(A'+6B'X)+12\kappa BHX}{(A+6BX)\sqrt{AX+3BX^2+U}}\, .
\ee

%\vspace{0.4cm}
The coefficients $A(\phi)$ and $B(\phi)$ are determined by the form of the potential function $V(\phi)$, which will be addressed in the following section. By substituting these coefficients and the real-valued solution for $\dot{\phi}[\phi]$ into the SRPs \eqref{eps1}-\eqref{eps4} above, we can derive the observational parameters as functions of the field $\phi$.

%%%%%%%%%%%%%%%%%%%%%%%%%%%%%%%%%%%%%%%%%%%%%%%%%%%%%%%%%%%%%%%%%%%%%%%%%%%%%%%%%%%%%%%%%%%%

\section{\texorpdfstring{$\beta$}{beta}-exponential model} \label{Sec.4}

In this section, we present the $\beta$-exponential model \cite{Martin:2013tda}. This potential was initially studied phenomenologically in \cite{Alcaniz:2006nu} as a generalization of power-law inflation, which has a potential proportional to the standard exponential function, i.e.,
\be \label{powexp}
  V(\phi) = V_0 \: \exp{(-\lambda \phi)}\; ,
\ee
and is investigated in \cite{Lucchin:1984yf}.

This pure exponential potential is modified using a deformation parameter $\beta$. As a result, the following generalized form for $V(\phi)$ is defined:
\be\label{genpot}
V(\phi) =  V_0 \: \exp_{1-\beta}{(-\lambda \phi )}\; ,
\ee
where the generalized $\beta$-exponential function $\exp_{1-\beta}$ is used. The definition of this function is as follows:
\begin{eqnarray} \label{genexp}
\mbox{$\exp_{1-\beta}{(x)}= $}   \; \left\{\begin{tabular}{l}
                                    $\left[1 + \beta x \right]^{1/\beta}\;  , \: 1 + \beta x > 0$\\
                                        \\
                                    $0$, \: otherwise.
                                \end{tabular}
                           \right.   
\end{eqnarray}
This generalized function satisfies, $\forall$ $ (x,\;y) < 0$, the following identities

\begin{equation}
\label{p1}
\exp_{1-\beta}\left[\ln_{1-\beta}({x}) \right] =  {x}\;, 
\end{equation}
and
\begin{equation}
\label{p2}
\ln_{1-\beta}({x}) + \ln_{1-\beta}({y}) =   \ln_{1-\beta}({xy}) -  \beta \left[\ln_{1-\beta}({x}) \ln_{1-\beta}({y})\right], 
\end{equation}
where $\ln_{1-\beta}({x}) = (x^{\beta} - 1)/\beta$ is the generalized logarithmic function.

Using the definition provided in equation \eqref{genexp}, we can write the explicit form of the $\beta$-exponential potential, as given in equation \eqref{genpot}. This results in the following expression:

\begin{equation} \label{jframebetamodel}
   V(\phi) = V_0\left[1 + \beta \left(-\lambda \phi \right) \right]^{1/\beta}\;.
\end{equation}

In this equation, the parameter $\beta$  determines the degree of interpolation between pure exponential (power-law) inflation and a wider range of behaviors. By adjusting the value of $\beta$, one can generate potentials that are effectively steeper or flatter than a pure exponential form. The variable $\lambda$ is a dimensionless parameter. This added flexibility leads to variations in the slow-roll parameters, which in turn affect the observables such as the spectral index $n_s$ and the tensor-to-scalar ratio $r$. Consequently, the model is able to fit the constraints from the CMB for various ranges of $\beta$ and $\lambda$. 

While the $\beta$-exponential potential was initially introduced phenomenologically as a generalized framework, it is deeply rooted in several well-motivated high-energy physics and thermodynamic contexts. Firstly, the generalized exponential function introduced in \eqref{genexp} is mathematically equivalent to the $q$-exponential function, $e_q(x) = [1 + (1-q)x]^{1/(1-q)}$, which emerges naturally from the maximization of Tsallis non-extensive entropy when $\beta = 1-q$ \cite{Tsallis1988, Borges:2004, Abe:2001}. In standard Boltzmann-Gibbs thermodynamics, entropy of the universe scales linearly with the horizon area ($S \propto A$), leading to logarithmic integrals during scalar field reconstruction that yield pure exponential potentials ($e^{-\lambda \phi}$). However, when non-extensive Tsallis thermodynamics—stemming from the long-range nature of gravitational interactions—is considered, the entropy becomes proportional to a fractional power of the area ($S \propto A^\delta$). 

The translation from such a non-extensive thermodynamic framework into an effective scalar field action is a well-established procedure known as scalar field reconstruction or THDE (Tsallis Holographic Dark Energy) correspondence \cite{Kumar2022}. By equating the effective density and pressure derived from modified Friedmann equations to a canonical scalar field ($\rho_\phi = \frac{1}{2}\dot{\phi}^2 + V$ and $p_\phi = \frac{1}{2}\dot{\phi}^2 - V$), the integration of these fractional powers naturally leads to $q$-logarithms. The inversion of these $q$-logarithms exactly generates the deformed $\beta$-exponential potentials. Thus, the potential in equation \eqref{jframebetamodel} is an analytical reflection of the early universe's non-extensive thermodynamic structure within the framework of quantum field theory \cite{Sheykhi2018, Nojiri2019}.

Secondly, from the perspective of extra-dimensional theories, the $\beta$-exponential form is rigorously grounded in braneworld cosmologies \cite{Santos:2017alg}. In this scenario, visible universe is a four-dimensional 3-brane embedded in a five-dimensional bulk, where the size of the extra dimension is governed by the radion field. By integrating out the five-dimensional action and applying the background BPS (Bogomol'nyi-Prasad-Sommerfield) solutions, the geometrical stabilization of the radion can be mapped onto an effective four-dimensional theory. The crucial physical mechanism occurs during the field redefinition mapping, where the inter-brane distance is transformed into a canonically normalized scalar field $\phi$. This mathematical mapping precisely converts the radion dynamics into the generalized $\beta$-exponential potential form. Consequently, the potential utilized in \eqref{jframebetamodel} provides a physically motivated, effective description that seamlessly complements the modified quadratic gravity approach treated in the Palatini formalism of \S \ref{Sec.2}.

The subsequent section will focus on the numerical analysis of how the various parameters of the model influence the observational indices.

%%%%%%%%%%%%%%%%%%%%%%%%%%%%%%%%%%%%%%%%%%%%%%%%%%%%%%%%%%%%%%%%%%%%%%%%%%%%%%%%%%

\section{Inflationary predictions} \label{Sec.5}

In this section, we present the numerical results for constant-roll $\beta$-exponential inflation within the Palatini formalism. The specific form of the inflaton potential in the Jordan frame, which drives the expansion of the universe, is given by equation \eqref{jframebetamodel}. In the Einstein frame, it takes the effective form $U(\phi)$ as stated in equation \eqref{abu}.

\begin{figure}[!htbp]%[H]
    \centering
    \includegraphics[width=\textwidth]{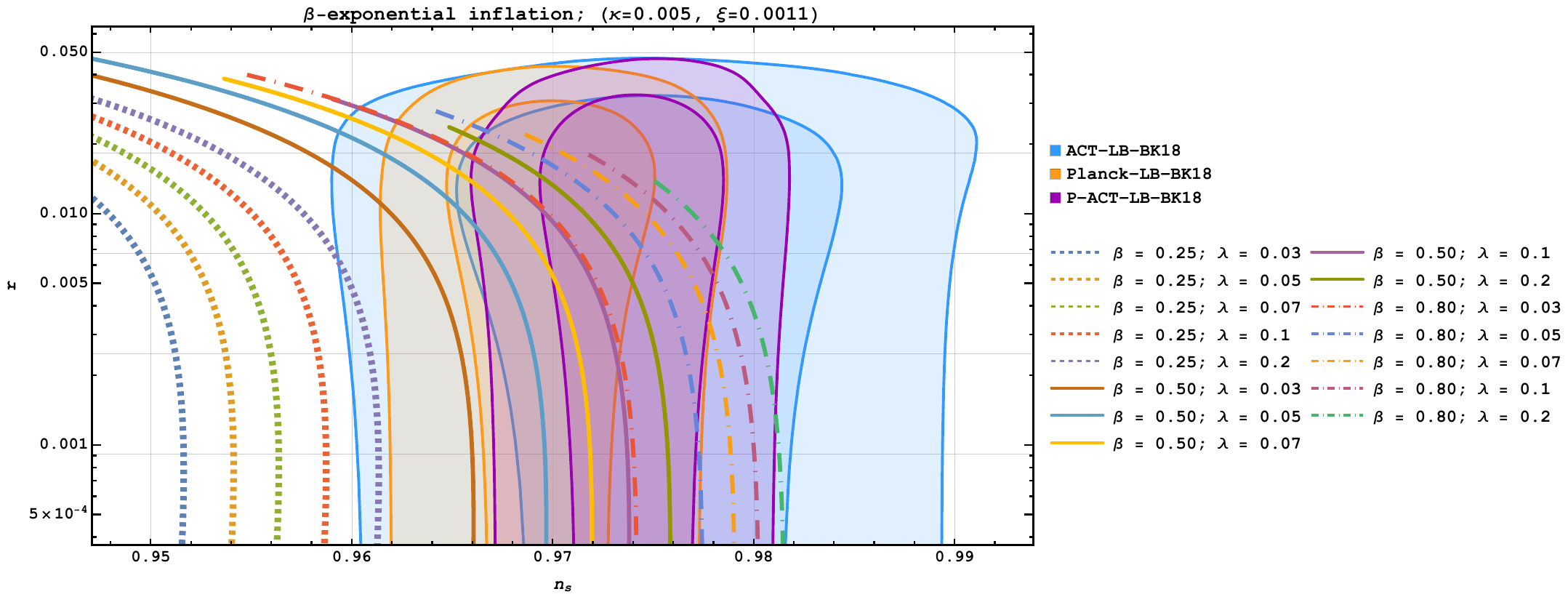}
    \caption{The predictions for the $n_s - r$ parameter space corresponding to selected values of the $\beta$ and $\lambda$ parameters for constant-roll $\beta$-exponential inflation within the Palatini formalism. The constant-roll parameter is fixed at $\kappa = 0.005$, and the non-minimal coupling is set at $\xi = 0.0011$. Meanwhile, the parameter $\alpha$ varies within the range of $[1 \times 10^5 - 8 \times 10^{10}]$. The light and dark contours represent the $95\%$ and $68\%$ confidence levels, respectively.
    The constraints on $r$ are driven by the {\textsf{BK18}} data \cite{BICEP:2021xfz}, while the constraints on $n_s$ are driven by {\textsf{Planck}} (orange) \cite{Akrami2018}, {\textsf{ACT}} (blue), or {\textsf{P-ACT}} (purple) \cite{ACTcollab2}. The combined dataset also includes CMB lensing and BAO in all cases.}
    \label{fig:fig1}
\end{figure}

Our numerical analysis of the inflationary parameters employs two distinct methods for calculating the e-folding parameter, $N_*$. To account for the evolution of the universe, we employ the instant reheating assumption in conjunction with the standard slow-roll (SR) method, as detailed in equation \eqref{perturb2}. Under the instant reheating assumption, $N_*$ is given by the following formula:
\begin{eqnarray} \label{efoldsreal2} 
N_*\approx64.7+\frac12\ln\rho_*-\frac{1}{4}\ln\rho_e \:.
\end{eqnarray}
The definition of $\rho_{*}$ is provided by:
\begin{equation}
\rho_{*} = \frac{3\pi^2 P_S \: r}{2},
\end{equation}
where $r$ is the tensor-to-scalar ratio. Furthermore, at the pivot scale, the energy density is approximately $\rho_{*} \approx U(\phi_{*})$, and the energy density at the end of inflation is $\rho_{e} = (3/2) U(\phi_{e})$. The end of the inflationary phase is dynamically determined by the standard graceful exit condition, where the first slow-roll parameter reaches unity ($\epsilon_1(\phi_e) = 1$). The evolution of the scalar field naturally guarantees this exit; as the field rolls down the effective potential, the kinetic contributions inherently grow to satisfy this condition, ensuring a smooth transition toward the subsequent reheating phase without requiring any ad hoc mechanisms.

\begin{figure}[htbp]
\centering
\includegraphics[width=\textwidth]{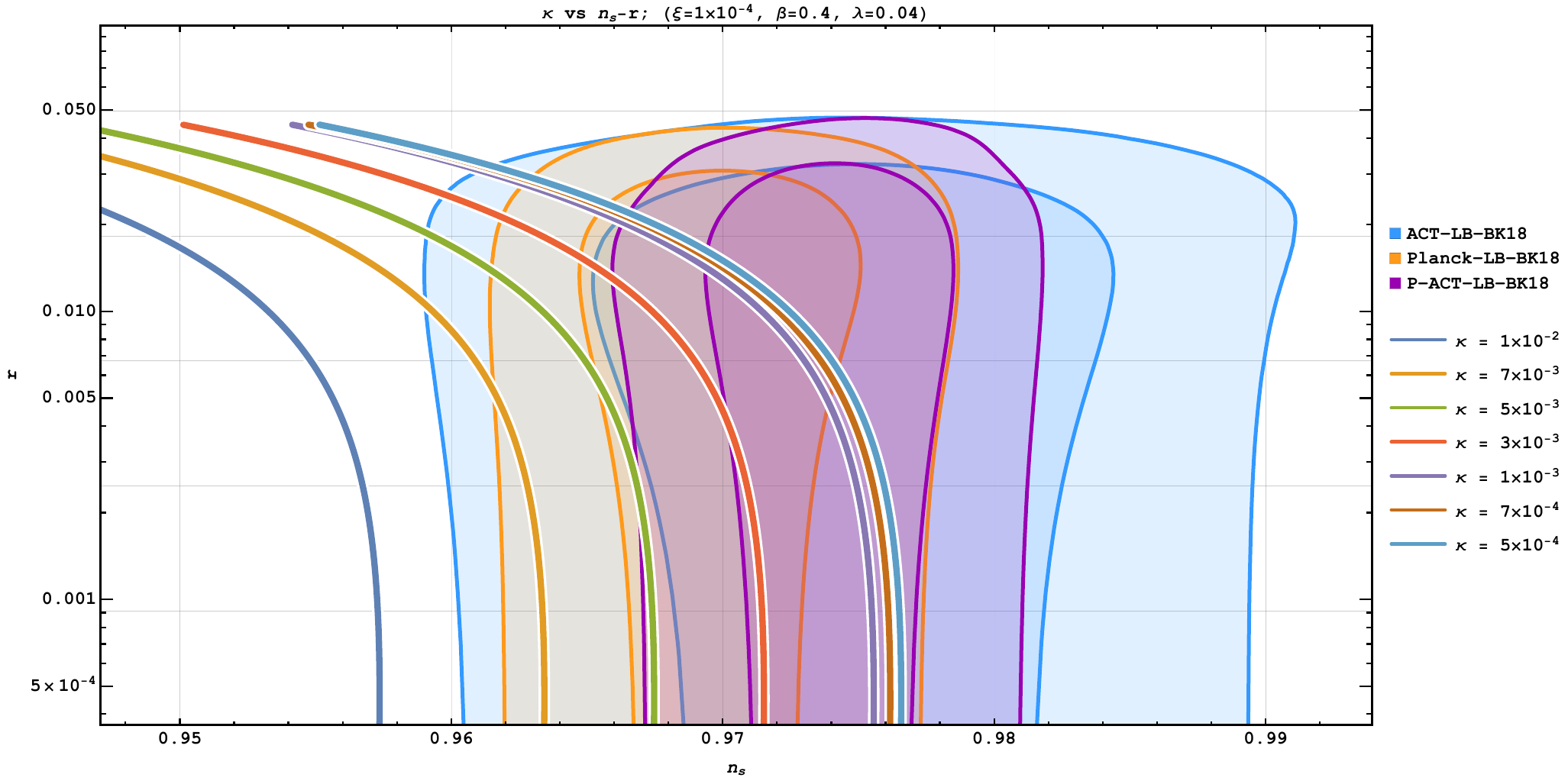}
\caption{The impact of the constant-roll parameter $\kappa$ on the $n_{s}-r$ parameter space within the framework of constant-roll $\beta$-exponential inflation in the Palatini formalism. The trajectories are plotted for distinct values of $\kappa$ ranging from $5\times10^{-4}$ to $1\times10^{-2}$, while the other model parameters are explicitly fixed at $\beta = 0.4$, $\lambda = 0.04$, and $\xi = 1\times10^{-4}$. The Starobinsky parameter $\alpha$ varies continuously over the range $[1\times10^{5}-8\times10^{10}]$, driving the downward trend toward lower tensor-to-scalar ratios. The light and dark shaded contours correspond to the 95\% and 68\% confidence levels, respectively, for the ACT-LB-BK18 (blue), Planck-LB-BK18 (orange), and P-ACT-LB-BK18 (purple) datasets.}
\label{fig:fig2}
\end{figure}

The inflationary parameters $n_s$ and $r$ are computed using an iterative algorithm that is primarily based on the parameter $V_0$ in the potential expressed in equation \eqref{jframebetamodel}. Finer iterations are subsequently carried out over the other variables of the model until the two different methods for computing the e-folding number converge. For the optimal benchmark values that center our predictions within the Planck and ACT contours, this algorithm yields an e-folding number of $N_{*} \approx 60$. While we employed the instant reheating approximation to derive these benchmark predictions, a prolonged reheating phase with a different equation of state could slightly shift this necessary number of e-folds, providing an additional physical mechanism to fine-tune the spectral index within the strict observational bounds.

In standard slow-roll inflation, observables such as the spectral index ($n_s$) and the tensor-to-scalar ratio ($r$) are rigidly tied to the shape of the scalar potential, creating a predictive degeneracy where different models yield indistinguishable results along a single, constrained curve. By introducing the constant-roll parameter ($\kappa$) alongside the Palatini non-minimal coupling ($\xi$), our model gains a new, independent degree of freedom that dynamically breaks this restrictive slow-roll lock, allowing the predicted curves to shift across the parameter space independently of the potential's geometry. The selected benchmark values, such as $\xi=0.0011$ and $\kappa=0.005$, are explicitly chosen as 'optimal' because they are large enough to successfully break this degeneracy, yet finely tuned enough to center the model's predictions exactly within the stringent $1\sigma$ confidence contours of the recent ACT DR6 and Planck datasets, thereby avoiding the excessive theoretical shifts that larger values would otherwise induce.

\begin{figure}[htbp]
\centering
 \includegraphics[width=\textwidth]{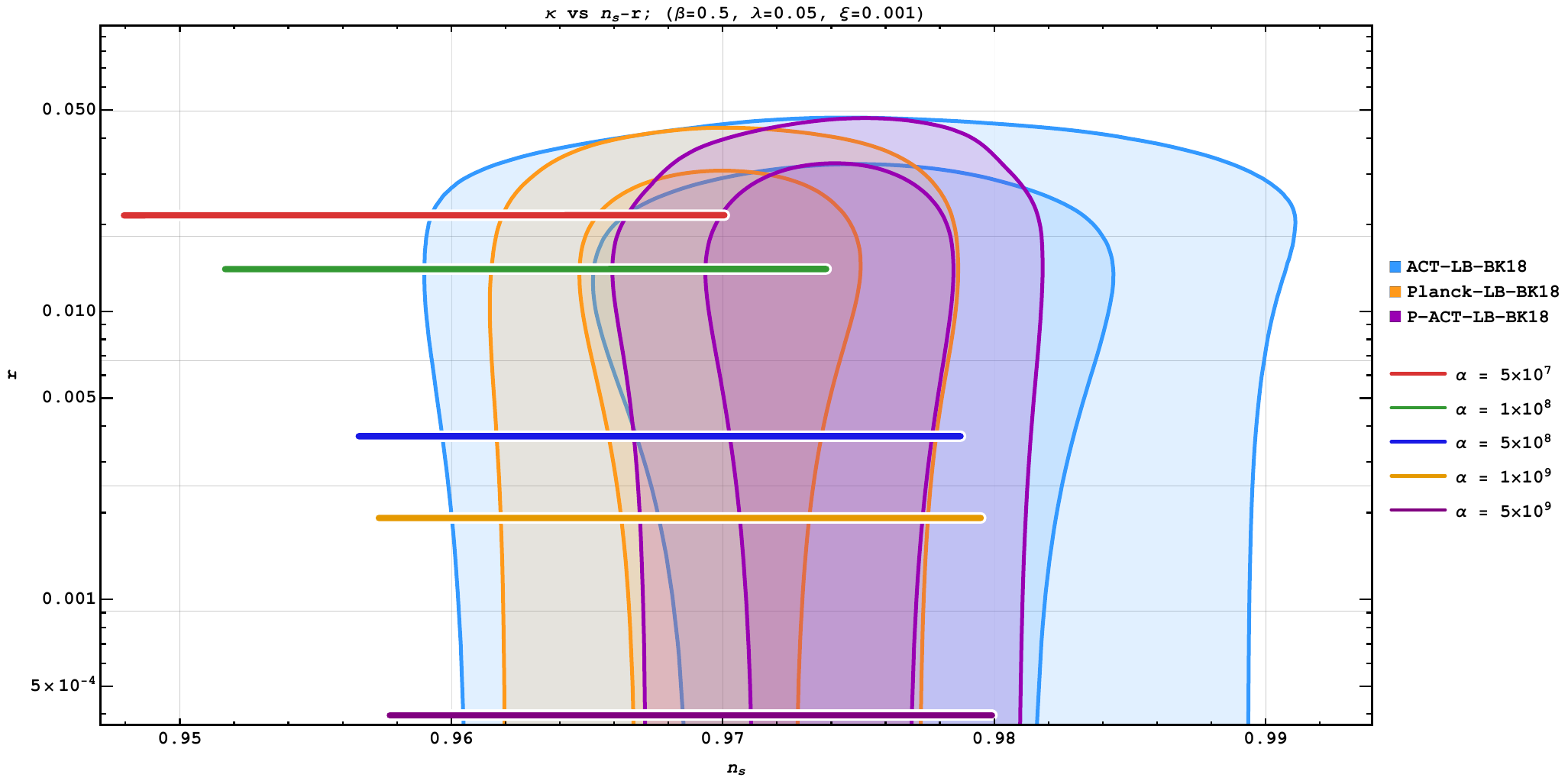}
\caption{The evolution of the inflationary predictions in the $n_{s}-r$ plane as the constant-roll parameter $\kappa$ varies continuously. The trajectories correspond to selected discrete values of the Starobinsky parameter $\alpha$, while the other model parameters are explicitly fixed at $\beta = 0.5$, $\lambda = 0.05$, and the non-minimal coupling $\xi = 1\times10^{-3}$. For each plotted curve, $\kappa$ increases continuously from $1\times 10^{-5}$ at the rightmost edge to $1.1\times 10^{-2}$ at the leftmost edge, demonstrating that larger constant-roll rates drive the spectral index $n_s$ strictly to lower values. The different curves from top to bottom illustrate the effect of increasing $\alpha$, which primarily suppresses the tensor-to-scalar ratio $r$. The light and dark shaded regions denote the 95\% and 68\% confidence intervals, respectively, for the ACT-LB-BK18 (blue), Planck-LB-BK18 (orange), and P-ACT-LB-BK18 (purple) datasets.}
\label{fig:fig3}
\end{figure}

Figure \ref{fig:fig1} illustrates the parameter space of $n_s - r$ for selected values of the constants $\beta$ and $\lambda$ in the context of constant-roll $\beta$-exponential inflation. In this plot, the constant-roll parameter is fixed at $\kappa=0.005$, and the non-minimal coupling is set to $\xi=0.0011$. Meanwhile, the Starobinsky parameter varies within the range of $\alpha \in [1\times 10^5 - 8\times 10^{10}]$. The constraints on $r$ are driven by the {\textsf{BK18}} data \cite{BICEP:2021xfz}, while the constraints on $n_s$ are driven by {\textsf{Planck}} (orange) \cite{Akrami2018}, {\textsf{ACT}} (blue), or {\textsf{P-ACT}} (purple) \cite{ACTcollab2}. The combined dataset also includes CMB lensing and BAO in all cases.

\begin{figure}[htbp]
\centering
\includegraphics[width=\textwidth]{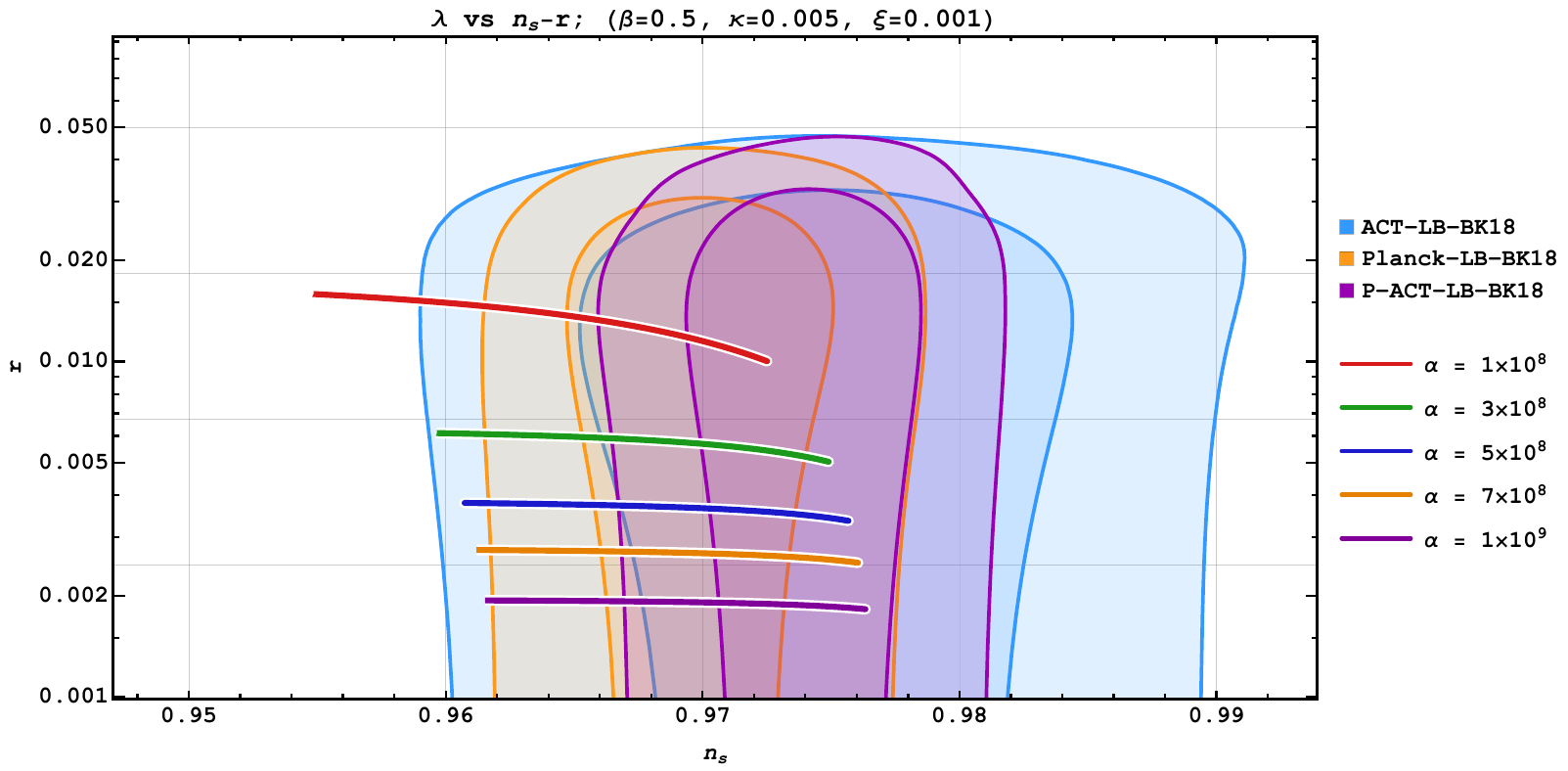}
\caption{The impact of the $\beta$-exponential parameter $\lambda$ on the inflationary predictions in the $n_{s}-r$ plane. The trajectories are plotted for selected discrete values of the Starobinsky parameter $\alpha$, while the other model parameters are explicitly fixed at $\beta = 0.5$, the constant-roll parameter $\kappa = 0.005$, and the non-minimal coupling $\xi = 1\times10^{-3}$. For each curve, $\lambda$ varies continuously from $0.001$ at the leftmost edge to $1.0$ at the rightmost edge, illustrating that an increase in $\lambda$ directly drives the spectral index $n_s$ toward larger values. Similar to the previous analyses, increasing $\alpha$ shifts the trajectories vertically downwards, systematically suppressing the tensor-to-scalar ratio $r$. The light and dark shaded regions represent the 95\% and 68\% confidence intervals, respectively, for the ACT-LB-BK18 (blue), Planck-LB-BK18 (orange), and P-ACT-LB-BK18 (purple) datasets.}
\label{fig:fig4}
\end{figure}

A comprehensive review of Figure \ref{fig:fig1} unveils the qualitative bounds required for the model's viability. Each curve corresponding to a specific ($\beta, \lambda$) pair begins with $\alpha = 1\times 10^5$ at the upper extreme and ends with $\alpha = 8\times 10^{10}$ at the lower end. This exhibits a striking characteristic: as the Starobinsky parameter escalates, the tensor-to-scalar ratio shows a notable decline. Specifically, to suppress the tensor-to-scalar ratio sufficiently to align with the {\textsf{BK18}} and {\textsf{ACT}} DR6 bounds, the Starobinsky parameter must reside in the qualitative range of $\alpha \gtrsim \mathcal{O}(10^8)$.

\begin{figure}[htbp]
\centering
\includegraphics[width=\textwidth]{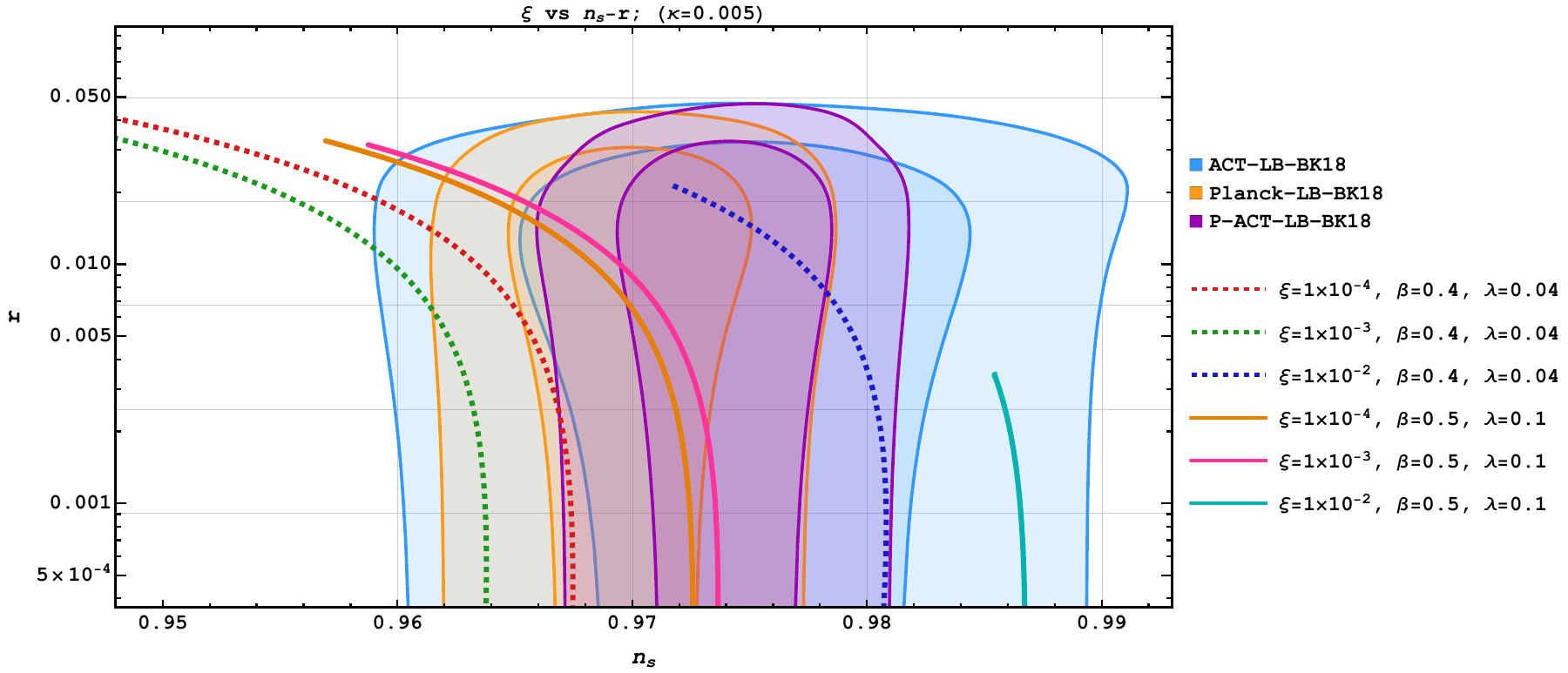}
\caption{The effect of the non-minimal coupling parameter $\xi$ on the inflationary predictions in the $n_s - r$ plane. The trajectories are plotted for selected $\xi$ values ($\xi = 1\times 10^{-4}, 1\times 10^{-3}, 1\times 10^{-2}$) across two distinct parameter sets: ($\beta = 0.4, \lambda = 0.04$) represented by dotted lines, and ($\beta = 0.5, \lambda = 0.1$) represented by solid lines. The constant-roll parameter is explicitly fixed at $\kappa = 0.005$ for all curves. Along each individual curve, the Starobinsky parameter $\alpha$ varies continuously within the range $[1\times 10^5 - 8\times 10^{10}]$, driving the predictions vertically toward lower tensor-to-scalar ratios $r$. Comparing curves with the same $(\beta, \lambda)$ configuration reveals that decreasing the non-minimal coupling $\xi$ systematically reduces the spectral index $n_s$, shifting the trajectories to the left. The shaded regions indicate the 95\% (light) and 68\% (dark) confidence levels for the ACT-LB-BK18 (blue), Planck-LB-BK18 (orange), and P-ACT-LB-BK18 (purple) datasets.}
\label{fig:fig5}
\end{figure}

To unequivocally elucidate the individual influence of the parameter $\beta$ in Figure \ref{fig:fig1}, a comparison of the curves corresponding to different $\beta$ values while holding $\lambda$ constant provides valuable insights. For instance, analyzing the three curves where $\lambda = 0.07$, as $\beta$ increases from 0.25 (green dashed line) to 0.8 (orange dotted-dashed line), the curves exhibit a pronounced shift to the right. Consequently, to prevent anomalous shifts and maintain the spectral index within the $1\sigma$ CLs of {\textsf{Planck}} and {\textsf{ACT}}, the deformation parameter is qualitatively constrained to the range $0.3 \lesssim \beta \lesssim 0.6$. A similar rightward shift is observed when $\lambda$ is increased while keeping $\beta$ fixed, underscoring the interconnectedness of these parameters.

To explicitly demonstrate the phenomenological advantage of our generalized framework, it is highly instructive to compare our numerical predictions with the minimal slow-roll case ($\xi=0, \kappa=0$) analyzed in Ref. \cite{bostan}. As demonstrated in Figure 3 of Ref. \cite{bostan}, the minimally coupled $\beta$-exponential model under the instant reheating assumption rigidly predicts a spectral index of $n_s \approx 0.966$ for the parameter set ($\beta=0.5, \lambda=0.1$). While this value is well-suited for the standard Planck 2018 constraints, the recent {\textsf{ACT}} DR6 and combined {\textsf{P-ACT}} datasets reveal a preference for a higher scalar spectral index ($n_s \gtrsim 0.970$). The minimal slow-roll model lacks the kinematic freedom to horizontally shift its predictions to this newly favored region without fundamentally altering the shape parameters or invoking prolonged reheating phases. 

In stark contrast, our analysis shows that extending the dynamics beyond the minimal slow-roll limit provides the necessary theoretical leverage to address this {\textsf{ACT}}-driven shift. As seen in our Figure \ref{fig:fig1}, the introduction of the constant-roll parameter ($\kappa = 0.005$) and the non-minimal coupling ($\xi = 0.0011$) dynamically shifts the spectral predictions to higher values. For instance, the parameter configuration ($\beta=0.5, \lambda=0.07$) elegantly centers the spectral index at an optimal $n_s \approx 0.972$, planting it directly within the stringent $1\sigma$ core of the {\textsf{P-ACT}} and {\textsf{ACT}} contours, while systematically suppressing the tensor-to-scalar ratio to $r \lesssim 0.005$ for $\alpha \gtrsim \mathcal{O}(10^8)$. This flawless alignment with the latest {\textsf{ACT}} constraints clearly illustrates that $\kappa$ and $\xi$ are essential physical features required to achieve optimal observational viability in light of the latest cosmological data.

To further underscore the theoretical necessity of the quadratic $R^2$ term and the Palatini approach, we can contrast our findings with the metric slow-roll analysis of the non-minimally coupled $\beta$-exponential model presented in Ref. \cite{dosSantos:2021vis}. As depicted in Figure 3 of Ref. \cite{dosSantos:2021vis}, which evaluates the predictions against the older {\textsf{Planck}} 2018 contours, the standard metric formalism struggles with an inherently large tensor-to-scalar ratio. Specifically, assuming the standard $N_* = 60$ e-folds (indicated by circular markers in their analysis) and a comparable non-minimal coupling of $\xi=0.001$, their metric model yields $r \gtrsim 0.02$ for deformation parameters $\beta \le 1.6$. For $\beta=0.8$, the predictions soar as high as $r \sim 0.06$, severely violating the latest stringent BICEP/Keck (BK18) \cite{BICEP:2021xfz} upper bound of $r < 0.036$. Consequently, the metric model is forced to invoke exceptionally large deformation parameters ($\beta \ge 2.5$) merely to suppress $r$ to acceptable levels.

In sharp contrast, our generalized Palatini framework introduces the Starobinsky parameter $\alpha$, which fundamentally modifies the kinetic structure of the inflaton. As evidenced by our parameter space scans in Figure \ref{fig:fig1}, the inclusion of the $R^2$ term acts as a powerful suppressor; increasing $\alpha \gtrsim \mathcal{O}(10^8)$ systematically drives the tensor-to-scalar ratio vertically down to $r \lesssim 0.005$, comfortably satisfying all current observational limits. Furthermore, this dynamic allows our framework to achieve excellent observational viability with much smaller, natural values of the shape parameter ($0.3 \lesssim \beta \lesssim 0.6$), as the synergistic effect of the constant-roll parameter $\kappa$ precisely calibrates the spectral index $n_s$ within the updated {\textsf{ACT}} DR6 limits. This direct comparison unequivocally demonstrates that the Palatini $R^2$ extension is not a mere parametric complication, but a vital physical requirement to rescue the model from the high tensor-to-scalar ratio tension inherent in the standard metric approach. 

This phenomenological advantage is further highlighted when comparing our results with the very recent metric slow-roll analysis of the $\beta$-exponential potential in Ref. \cite{Yuennan2026}. As depicted in Figures 1 and 4 of Ref. \cite{Yuennan2026}, while the metric slow-roll model successfully elevates the spectral index into the {\textsf{ACT}}-favored region ($n_s \approx 0.974$), the tensor-to-scalar ratio remains persistently bounded around $r \sim \mathcal{O}(10^{-2})$. Even with the inclusion of non-minimal coupling ($\xi \sim 10^{-4}$), their framework requires large deformation parameters ($\beta \sim 5.0$) to suppress tensor modes to $r \approx 0.01$. In stark contrast, our generalized Palatini constant-roll framework achieves a much more rigorous suppression of the tensor modes ($r \lesssim 0.005$) via the $\alpha R^2$ term, while seamlessly calibrating $n_s$ with the constant-roll parameter $\kappa$. Consequently, our model flawlessly satisfies both the {\textsf{ACT}} DR6 and the latest BICEP/Keck constraints while utilizing significantly smaller, mathematically natural values for the deformation parameter ($0.3 \lesssim \beta \lesssim 0.6$).

The effect of the constant-roll parameter is examined in Figure \ref{fig:fig2} and Figure \ref{fig:fig3}. Figure \ref{fig:fig2} illustrates the $n_s - r$ curves for chosen values of $\kappa$, while keeping $\beta$, $\lambda$, and $\xi$ fixed. Higher values of $\kappa$ translate the curves to the left, moving them toward the region of lower spectral index. For example, the red curve with $\kappa = 3 \times 10^{-3}$ bisects the Planck 68\% confidence level contour right in the center. In contrast, the dark blue curve with $\kappa = 1 \times 10^{-2}$ falls completely outside all viable contours. This establishes a strict qualitative bound for the constant-roll parameter at $\kappa \sim \mathcal{O}(10^{-3})$, demonstrating that even minor deviations toward $\mathcal{O}(10^{-2})$ completely rule out the model. Figure \ref{fig:fig3} enhances this analysis by illustrating the $n_s - r$ curves where $\kappa$ varies continuously from $1\times 10^{-5}$ to $1.1\times 10^{-2}$ for specific values of $\alpha$, confirming that as $\kappa$ increases, the spectral index decreases while $r$ remains relatively stable.

The subsequent plot, Figure \ref{fig:fig4}, investigates how the observational indices vary continuously with the $\beta$-exponential parameter $\lambda$ in the range of $[0.001 - 1.0]$. For each curve corresponding to a specific $\alpha$, as $\lambda$ increases, the spectral index $n_s$ also tends to increase (e.g., for $\alpha = 7\times 10^8$, $n_s$ shifts from $0.961$ to $0.976$).

The final plot of the analysis, Figure \ref{fig:fig5}, demonstrates the variation in response to the alterations in the non-minimal coupling parameter $\xi$. In general, decreasing the non-minimal coupling reduces the spectral index, shifting the predictions to the left. To optimally center the theoretical predictions within the observational contours, the non-minimal coupling must be fine-tuned around the suitable range of $\xi \sim \mathcal{O}(10^{-3})$.

Furthermore, the constant-roll dynamics within this Palatini generalized k-inflation framework yield distinct phenomenological signatures regarding primordial non-Gaussianities, strictly differentiating it from standard canonical slow-roll scenarios. To substantiate the physical significance of the constant-roll condition, we evaluate the equilateral non-Gaussianity amplitude $f_{NL}^{equil}$. Following the general formalism for constant-roll k-inflation dynamics detailed in Refs. \cite{Odintsov2020} and \cite{Felice2011},the dominant contributions can be parameterized using the sound speed $C_s^2$ and the slow-roll parameters. For our specific effective Lagrangian $\mathcal{L}(\phi, X) = A(\phi)X + B(\phi)X^2 - U(\phi)$, the higher-order kinetic derivative vanishes ($\mathcal{L}_{XXX} = 0$). By expanding the non-Gaussianity amplitude to the leading order in the slow-roll parameters and utilizing the constant-roll relation $\epsilon_2 = -\kappa$, the amplitude takes the form:
\begin{equation} \label{nongauss}
f_{NL}^{equil} \approx \frac{85}{81} \left( \frac{1}{C_s^2} - 1 \right) - \frac{10}{81} \left( \frac{1}{C_s^2} - 1 \right) \left( \frac{2B(\phi)X^2}{A(\phi)X + 6B(\phi)X^2} \right) + \mathcal{C}_1 \epsilon_1 - \mathcal{C}_2 \kappa + \mathcal{C}_3 s
\end{equation}
where $\mathcal{C}_1, \mathcal{C}_2$, and $\mathcal{C}_3$ are $\mathcal{O}(1)$ theoretical coefficients, and $s \equiv \dot{C}_s / (H C_s)$ represents the running of the sound speed. 

Evaluating the dominant zeroth-order terms of Eq. \eqref{nongauss} numerically at horizon crossing ($N_* \approx 60$) utilizing our benchmark parameters yields a non-vanishing baseline amplitude of $f_{NL}^{equil} \approx +0.006$. In standard metric slow-roll models, the sound speed is canonical ($C_s^2 = 1$) and $\kappa \to 0$, reducing the amplitude to a vanishingly small quantity governed solely by the first slow-roll parameter ($\sim \mathcal{O}(\epsilon_1)$). Conversely, in our Palatini generalized framework, the field-dependence of $A(\phi)$ and $B(\phi)$ dynamically enforces a sub-luminal sound speed ($C_s^2 \approx 0.994$), inherently generating this distinct positive baseline shift. Furthermore, beyond these sub-luminal sound speed corrections, the emergence of the $-\mathcal{C}_2 \kappa$ term introduces a direct, linear non-Gaussian correction driven entirely by the constant-roll rate. Crucially, given our optimal benchmark value of $\kappa \sim \mathcal{O}(10^{-3})$, the combined magnitude of these corrections remains theoretically distinct, yet sufficiently small to safely satisfy the current observational bounds set by the Planck collaboration ($f_{NL}^{equil} = -26 \pm 47$) \cite{Planck2019}. This clearly demonstrates that the constant-roll parameter $\kappa$ is not a mere background evolution auxiliary, but a fundamental physical quantity that directly modulates the primordial non-Gaussian fingerprints—leaving a unique phenomenological signature that strictly differentiates our model from standard slow-roll approximations while safely preserving observational viability.

%%%%%%%%%%%%%%%%%%%%%%%%%%%%%%%%%%%%%%%%%%%%%%%%%%%%%%%%%%%%%%%%%%%%%%%%%%%%%%%%%%
\section{Summary and Conclusions} \label{Sec.conc}

In this article, we have systematically investigated the inflationary dynamics of a non-minimally coupled $\beta$-exponential model within the Palatini formalism, incorporating a quadratic $R^{2}$ gravity term under the constant-roll condition. By deriving the equations of motion and analyzing the effective Einstein-frame Lagrangian, we established that the model corresponds to a generalized k-inflation theory. This framework accommodates a broad spectrum of physically motivated inflationary solutions that are in excellent agreement with the most recent constraints from the {\textsf{ACT}} DR6 and {\textsf{Planck}} missions.

A central finding of our analysis is that constant-roll dynamics offer a significantly richer and more adaptable phenomenological landscape than the standard slow-roll approximation. In standard slow-roll inflation, the predictive degeneracy rigidly ties the observables ($n_s$ and $r$) to the geometric shape of the scalar potential. However, our model demonstrates that the introduction of the constant-roll parameter ($\kappa$), alongside the Palatini non-minimal coupling ($\xi$), injects a new, independent degree of freedom. This mechanism dynamically breaks the restrictive slow-roll lock, providing the theoretical flexibility required to shift the predicted curves across the parameter space independently of the potential's slope.

To ensure strict alignment with the $1\sigma$ confidence levels of the combined {\textsf{ACT}} and {\textsf{Planck}} datasets, our parameter space scan identified specific qualitative and quantitative bounds for the model. Suppressing the tensor-to-scalar ratio below the stringent observational upper limits dictates a sufficiently large Starobinsky parameter, qualitatively bounded as $\alpha \gtrsim \mathcal{O}(10^{8})$. Furthermore, the spectral index is highly sensitive to the constant-roll rate; while a benchmark value of $\kappa = 5 \times 10^{-3}$ optimally centers the model within the viable regions, larger values such as $\kappa = 1 \times 10^{-2}$ induce excessive theoretical shifts, moving the predictions entirely out of the permitted bounds. Crucially, this optimized constant-roll regime also leaves a distinct imprint on the primordial bispectrum, generating a non-vanishing yet heavily suppressed equilateral non-Gaussianity amplitude ($f_{NL}^{equil} \approx +0.006$) that comfortably satisfies the current bounds set by the Planck collaboration. The non-minimal coupling and the deformation parameter further refine this fit, requiring an optimal calibration around $\xi \sim \mathcal{O}(10^{-3})$ and $0.3 \lesssim \beta \lesssim 0.6$, respectively.

Finally, our study highlights the distinctive advantages of the Palatini formulation over the standard metric approach. In the metric case, the presence of an $R^{2}$ term typically introduces a second scalar degree of freedom (the scalaron), effectively transforming the system into a two-field problem. In contrast, the Palatini formalism preserves the single-field nature of the inflation while fundamentally modifying the kinetic structure of the effective theory. This unique kinetic modification naturally leads to a suppressed tensor-to-scalar ratio that is more easily reconciled with current {\textsf{BICEP/Keck}} and {\textsf{ACT}} data. The theoretical robustness provided by the physical origins of the $\beta$-exponential potential, combined with the phenomenological flexibility of the parameters ($\beta, \lambda, \kappa, \xi, \alpha$), renders this model a compelling candidate for elucidating the complex dynamics of the early universe.

%%%%%%%%%%%%%%%%%%%%%%%%%%%%%%%%%%%%%%%%%%%%%%%%%%%%%%%%%%%%%%%%%%%%%%%%%%%%%%%%

\section*{Acknowledgment}

The author thanks Canan Karahan for the insightful discussions on the subject.


%\bibliographystyle{ptephy}
%\bibliography{sample}
%
% once the .bbl file has been generated then place the text in your article.

\vspace{0.2cm}
\noindent
%For references,  note how to include DOI information from examples below. 

%This is added by T. Yoneya (editor-in-chief) on 2020/07/09.

\let\doi\relax

%without this code before the command "\begin{thebibliography}{}" , an error will be %flagged. When the bibliography is provided as separate .bib file, then this code %should be placed above the commands "\bibliographystyle{}" and "\bibliography{}" %inside the main TeX file. 

\end{document}